\documentclass[aps,pra,reprint,superscriptaddress]{revtex4-2}
\usepackage{graphicx}
\usepackage{float}
\usepackage{blindtext}
\usepackage{physics}
\usepackage{gensymb}
\usepackage[table]{xcolor}
\usepackage{cancel}
\usepackage{mathtools, bm}
\usepackage{amssymb, bm}

\begin{document}

\title{Optimal Floquet Engineering for Large Scale Atom Interferometers}

\author{T. Rodzinka}
\affiliation{Laboratoire Collisions Agrégats Réactivité (LCAR/FERMI), UMR5589, Université
Toulouse III - Paul Sabatier and CNRS, 118 Route de Narbonne, F-31062 Toulouse, France}
\author{E. Dionis}
\affiliation{Laboratoire Interdisciplinaire Carnot de Bourgogne, CNRS UMR 6303, Université de Bourgogne, BP 47870, F-21078 Dijon, France}
\author{L. Calmels}
\affiliation{Laboratoire Collisions Agrégats Réactivité (LCAR/FERMI), UMR5589, Université
Toulouse III - Paul Sabatier and CNRS, 118 Route de Narbonne, F-31062 Toulouse,
France}
\author{S. Beldjoudi}
\affiliation{Laboratoire Collisions Agrégats Réactivité (LCAR/FERMI), UMR5589, Université
Toulouse III - Paul Sabatier and CNRS, 118 Route de Narbonne, F-31062 Toulouse,
France}
\author{A. Béguin}
\affiliation{Laboratoire Collisions Agrégats Réactivité (LCAR/FERMI), UMR5589, Université
Toulouse III - Paul Sabatier and CNRS, 118 Route de Narbonne, F-31062 Toulouse,
France}
\author{D. Guéry-Odelin}
\affiliation{Laboratoire Collisions Agrégats Réactivité (LCAR/FERMI), UMR5589, Université
Toulouse III - Paul Sabatier and CNRS, 118 Route de Narbonne, F-31062 Toulouse,
France}
\author{B. Allard}
\affiliation{Laboratoire Collisions Agrégats Réactivité (LCAR/FERMI), UMR5589, Université
Toulouse III - Paul Sabatier and CNRS, 118 Route de Narbonne, F-31062 Toulouse,
France}
\author{D. Sugny}
\affiliation{Laboratoire Interdisciplinaire Carnot de Bourgogne, CNRS UMR 6303, Université de Bourgogne, BP 47870, F-21078 Dijon, France}
\author{A. Gauguet}
\affiliation{Laboratoire Collisions Agrégats Réactivité (LCAR/FERMI), UMR5589, Université
Toulouse III - Paul Sabatier and CNRS, 118 Route de Narbonne, F-31062 Toulouse,
France}
\email{gauguet@irsamc.ups-tlse.fr}


\begin{abstract}
The effective control of atomic coherence with cold atoms has made atom interferometry an essential tool for quantum sensors and precision measurements. The performance of these interferometers is closely related to the operation of large wave packet separations. We present here a novel approach for atomic beam splitters based on the stroboscopic stabilization of quantum states in an accelerated optical lattice. The corresponding Floquet state is generated by optimal control protocols. In this way, we demonstrate an unprecedented Large Momentum Transfer (LMT) interferometer, with a momentum separation of 600 photon recoils ($600\hbar k$) between its two arms. Each LMT beam splitter is realized in a remarkably short time (2 ms) and is highly robust against the initial velocity dispersion of the wave packet and lattice depth fluctuations. Our study shows that Floquet engineering is a promising tool for exploring new frontiers in quantum physics at large scales, with applications in quantum sensing and testing fundamental physics.
\end{abstract}

\maketitle

Atom interferometry has made significant contributions to quantum technologies, enabling advances in inertial sensing~\cite{Bongs2019,Wu19,Geiger2020}, and the measurement of fundamental physical constants~\cite{Parker18,Morel20,Rosi2014}. It also has great potential for performing fundamental tests, such as testing the weak equivalence principle~\cite{Asenbaum2020}, searching for the nature of dark energy~\cite{Jaffe2017,Sabulsky19}, or investigating analogues of the Aharonov-Bohm effect~\cite{Gillot13,Overstreet22}. Enlarging the spatial separation between the arms of an interferometer holds great promise for increasing the sensitivity of quantum sensors. It is also instrumental in the exploration of new physics at the interface of relativity and quantum mechanics~\cite{Loriani2019,Overstreet2023,Carney2021}, as well as to the detection of gravitational waves~\cite{Dimopoulos2008,Canuel2018short,Badurina2020,Zhan2020,Schlippert2020,Abe2021short} and dark matter~\cite{Geraci2016,Arvanitaki2018}. These proposals highlight the critical need for highly efficient atom manipulation processes, especially to achieve momentum separations greater than 1000 photon recoils ($1000\hbar k$). Large Momentum Transfer (LMT) techniques increase the momentum separation from a superposition of two states generated by a $\pi/2$ beam splitter pulse. This low momentum separation is further enhanced by continuous acceleration via Bloch oscillations~\cite{Clade2009,McDonald2013,Pagel2020} or discrete acceleration using $\pi$ pulse sequences~\cite{Chiow2011,Plotkin2018,Rudolph2020,Beguin2023}. To date, the largest momentum transfer techniques used in interferometers have demonstrated momentum separations up to 400$\hbar k$~\cite{Gebbe2021,Wilkason2022}.

In this paper, we present a novel approach that unifies discrete and continuous acceleration methods by combining the Floquet formalism with quantum Optimal Control Theory (OCT). Quantum OCT is a set of methods for designing electromagnetic fields to perform specific quantum operations with optimal efficiency~\cite{PRXQuantumsugny}. It has emerged as a key tool in the advancement of quantum technologies~\cite{kochroadmap}. In the context of atom interferometry, OCT has only been successfully applied to a limited number of momentum states, e.g. to improve the robustness of interferometers based on Raman beam splitters~\cite{Saywell2020} or third-order Bragg diffraction~\cite{saywell2023}. However, the use of these protocols in optical lattice-based LMT experiments remains challenging due to the significant number of states involved and the need for robustness over a wide range of parameters~\cite{Goerz2023,Louie2023}. In our approach, optimal control protocols are employed to guarantee the robust preparation of Floquet states against the velocity dispersion of the atomic ensemble. This Floquet state based approach significantly reduces the complexity of the system under consideration. Consequently, it allows the application of OCT in situations where the control problem would be numerically intractable without this formalism.

This method operates in deep non-adiabatic regimes, enabling remarkably fast and highly effective acceleration within an optical lattice, exceeding the current state of the art. We demonstrate an interferometer capable of achieving a momentum separation of up to 600$\hbar k$ with a visibility of 20$\%$. In addition, numerical simulations confirm the scalability of our approach and suggest the possibility of atom interferometers larger than 1000$\hbar k$. This advance paves the way for significant progress in the measurement of the fine structure constant and addresses a critical challenge for atom interferometers operating at very large scales, particularly in the context of gravitational wave detectors.

\begin{figure*}
\centerline{\includegraphics[width=1.0\textwidth]{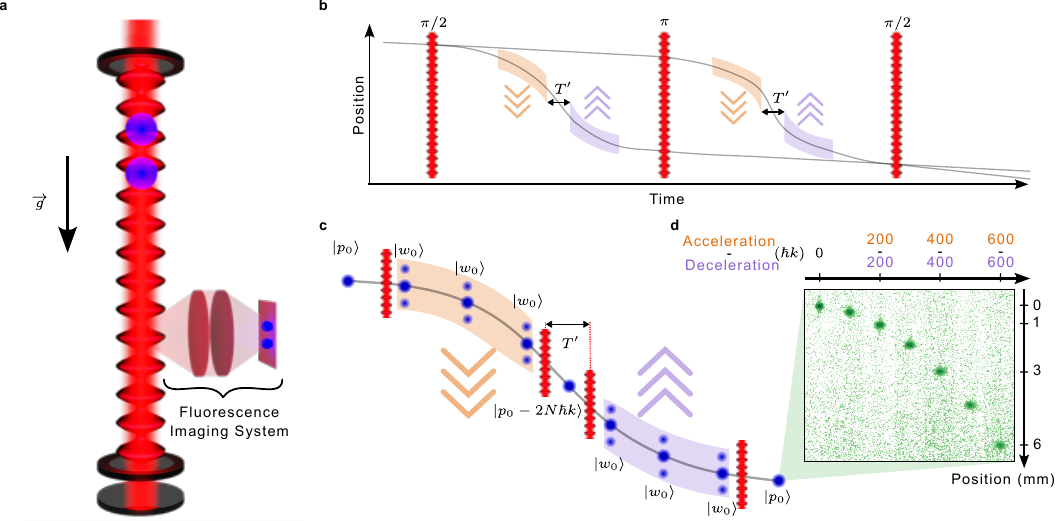}}
\caption{\textbf{Large Momentum Transfer Interferometer} 
(a)~Scheme of the experimental setup. A vertical optical lattice is used to manipulate the atom momentum states. Fluorescence imaging is used to detect the atoms after a time of flight.
(b)~Space-time diagram of the LMT interferometer based on a sequence consisting of two $\pi/2$ pulses separated by a $\pi$ pulse. Between these pulses, the two arms are successively accelerated and decelerated by a sequence of additional lattice pulses. The atoms are in free fall for $T'$ between the acceleration and deceleration stages. 
(c)~Scheme of the acceleration-deceleration process. Blue dots represent the momentum state decomposition at different times of the sequence. (d) Stack of images of atomic ensembles for different maximum momentum separations. The images are taken after a full sequence of acceleration and deceleration for a single arm and after a time of flight of 14~ms. A maximum separation of 600$\hbar k$ corresponds to a transfer of 1200$\hbar k$ per arm (acceleration + deceleration).
}
\label{fig:apparatus}
\end{figure*}

\subsection*{Experimental setup}
Our experimental setup, shown in Fig.~\ref{fig:apparatus}(a), uses a Bose-Einstein Condensate (BEC) consisting of about $3\times 10^4$ $^{87}\mathrm{Rb}$ atoms. After a free fall of about 5~ms, the atoms have an initial center-of-mass momentum of $\sim 8\hbar k$ in the laboratory frame and a momentum dispersion of $0.3\hbar k$ corresponding to an effective temperature of approximately 30~nK. The atoms then interact with a retroreflected vertical optical lattice, with tunable frequency and amplitude. The acceleration due to gravity is compensated by applying a linear frequency ramp to achieve a stationary lattice in the free fall frame. 

The experimental configuration implements an analog of the Mach-Zehnder interferometer for atoms by a sequence of optical lattice pulses inducing Bragg transitions between momentum states~\cite{Beguin2022}, as shown in Fig.~\ref{fig:apparatus}(b). The initial $\pi/2$ pulse creates a coherent superposition between two momentum states, $\ket{p_0}$ and $\ket{p_0 - 2\hbar k}$, thus acting as a beam splitter for the matter wave. Here, $p_0\ll \hbar k$ denotes the initial momentum of an atom in the free fall frame. The lower path then undergoes an acceleration sequence and is decelerated again after a free propagation time $T$'. This acceleration-deceleration sequence (Fig.~\ref{fig:apparatus}(c)) ideally does not affect the upper arm. Then a $\pi$ pulse reverses the momentum states $\ket{p_0}$ and $\ket{p_0 - 2\hbar k}$, acting as a mirror for the two arms. The same sequence of acceleration, free propagation $T^\prime$ and deceleration is then applied to the upper path. Finally, a $\pi/2$ pulse acts as a second beam splitter to complete the interferometer. The populations in the two main output ports with states $\ket{p_0}$ and $\ket{p_0-2 \hbar k}$ are measured through fluorescence imaging after $\sim 14$~ms of free fall time.

\subsection*{Principle of Floquet atom accelerator}
\begin{figure*}
\centerline{\includegraphics[width=1.0\textwidth]{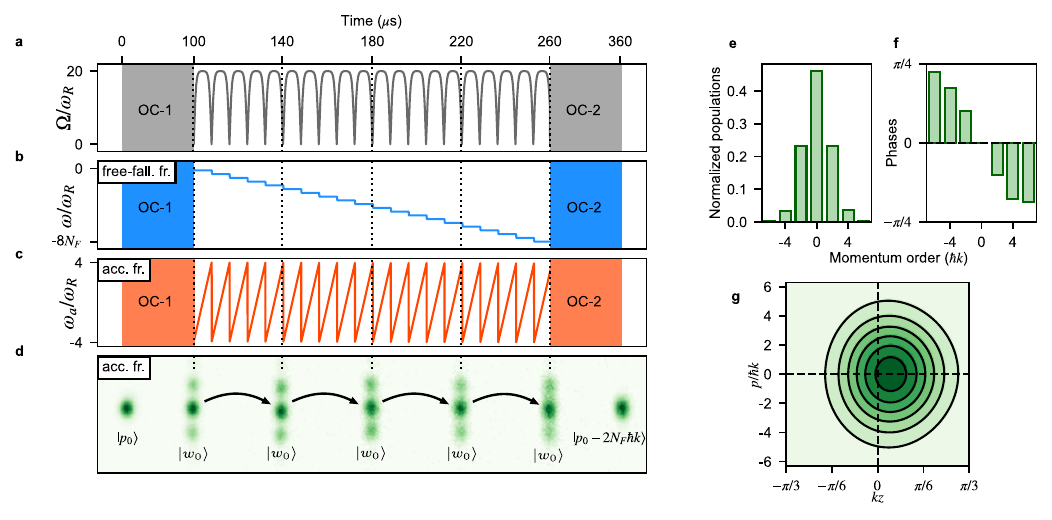}}
\caption{\textbf{Principle of Floquet acceleration.}
(a)~The time variation of the optical lattice consists of a periodic series of $\pi$ pulses of duration $\tau$ and amplitude $\Omega$, illustrated here with tanh pulses.
(b)~The resonance condition in the laboratory frame is adjusted for each pulse, resulting in a stepwise evolution of the lattice frequency. 
(c)~In the accelerated frame, the lattice frequency $\omega_a$ shows a periodic sawtooth shape, resulting in $\tau$-periodic driving of the lattice. OC-1 and OC-2 represent the optimal Floquet state preparation pulses. 
(d)~Stack of experimental images showing accelerated atoms at different steps of the acceleration sequence for $N_F=20$. The images are taken after a time-of-flight showing the momentum distributions of the input state $\ket{p_0}$, the output state $\ket{p_0-2N_F\hbar k}$ and the prepared Floquet state $\ket{w_0}$ during the periodic acceleration sequence.
(e)-(f)~Populations and phases of the transported Floquet state in the momentum states basis for a 5.3~$\mu$s square pulse. 
(g)~Corresponding phase space Husimi representation of the Floquet state (see Supp. Mat.). 
}
\label{fig:Strob-states}
\end{figure*}

The acceleration of the atom is achieved by a sequence of optical lattice $\pi$ pulses, each of duration $\tau$. Every pulse transfers a momentum of $2 \hbar k$ to the atom. This results in an average acceleration of $a_l = \frac{2 \hbar k}{M \tau}$, where $M$ is the atomic mass. The two-photon resonance condition is adjusted for each pulse to ensure efficient momentum transfer. The frequency difference $\omega$ between the two arms of the optical lattice, in the free fall frame, driving the Bragg transitions is therefore a piecewise constant function with a decrease of $8 \omega_r = 4 \frac{\hbar k^2}{M}$ between pulses, as represented in Fig.~\ref{fig:Strob-states}(b). We define the accelerated frame as the reference frame following the average acceleration $a_l$ of the atoms. In this frame, the lattice frequency is a periodic sawtooth function with period $\tau$, as shown in Fig.~\ref{fig:Strob-states}(c). In the accelerated frame, the Hamiltonian is periodic, $\hat{H}(t) = \hat{H}(t + \tau)$, and the dynamics of the system is naturally described within the Floquet formalism~\cite{Bukov2015}.

Floquet's theorem states that there exists a complete set of solutions to the time-dependent Schrödinger equation, called Floquet states, which can be obtained by diagonalizing the one-period propagator~\cite{Shirley1965} (see Supp. Mat.). Here the Floquet’s theorem simplifies as the system is observed stroboscopically at times $n\tau$. In particular, if the system is prepared in a single Floquet state $\ket{w_m (t_0)}$ at time $t_0$, its temporal evolution exhibits periodicity up to a phase factor:
\begin{equation}
    \ket{w_m(t_0+\tau)} = \ket{w_m(t_0)} e^{i\theta_m}\;.
\end{equation}
The principle of the Floquet accelerator is based on the stroboscopic stabilization of the system, i.e., the atom returns to the same Floquet state $\ket{w_m}$ after each pulse of duration $\tau$ (see Fig.~\ref{fig:Strob-states}(d)). It results in a lossless coherent acceleration of the atomic wave packet in the free fall frame. Among all the Floquet states, a relevant choice is the state $\ket{w_0}$, which is localized in both position and momentum within each lattice cell, and has the highest projection onto the initial momentum state $\ket{p_0}$. This specific Floquet state appears in particular for a $\pi$ pulse and depends on the temporal pulse shape. For sufficiently short pulse duration $\tau$, this state is very similar to a displaced and squeezed state in phase space~\cite{schleich2011}, which is preserved during the dynamics and whose time evolution of the position and momentum expectation values is well described by the corresponding classical trajectory (see Supp. Mat.). An example is shown in Fig.~\ref{fig:Strob-states}(g) with the Husimi representation of the Floquet state. The displacement in position of the quantum state results from a balance between the inertial force $-Ma_l$ in the accelerated frame and the restoring force due to the optical lattice.

The efficiency of the acceleration is greatly improved by adding a suitably shaped pulse to prepare the Floquet state. In practice, this pulse is designed using OCT to adjust both the amplitude and frequency of the optical lattice. This step is performed before (and at the end of) the acceleration sequence, and it transforms the initial state $\ket{p_0}$ into the corresponding Floquet state $\ket{w_0}$ defined for a given $\pi$ pulse of the acceleration sequence (and vice versa). The corresponding optimal control protocols are denoted OC-1 and OC-2 in Fig.~\ref{fig:Strob-states}. Similar sequences are used during the deceleration phase.

Figure~\ref{fig:Strob-states} shows the principle of Floquet acceleration with a sequence of $N_F=20$ pulses that transfers 40$\hbar k$. At each step of the sequence, a time-of-flight measurement allows the atomic states to be mapped onto the momentum state basis $\ket{p_0+2n\hbar k}$, with ${n\in\mathbb{Z}}$. In Fig.~\ref{fig:Strob-states}(d), the different states are displayed in the accelerated frame to improve the readability of the images. 

The Floquet acceleration relies on the periodicity of the pulse sequence, independent of the specific shape of each pulse. Therefore, the Floquet state $\ket{w_0}$ can be identified for various types of pulses, including sequential Bragg pulses, continuous Bloch-type accelerations (such as adjacent square pulses of constant frequency in the accelerated frame), and their combinations. In this study, we use discrete frequency evolution and either discrete (hyperbolic tangent tanh pulses) or continuous (adjacent square pulses) amplitude evolution. Detailed amplitude and frequency profiles of the lattice are given in the Methods section. 

It is worth noting that our approach differs from the use of Floquet's formalism in \cite{Wilkason2022}, where it is used to find periodic amplitude modulations within a single pulse to mitigate the opposite Doppler effects of the two arms of an interferometer.

\subsection*{Robust Floquet state preparation}
\begin{figure}
\centerline{\includegraphics[width=0.5\textwidth]{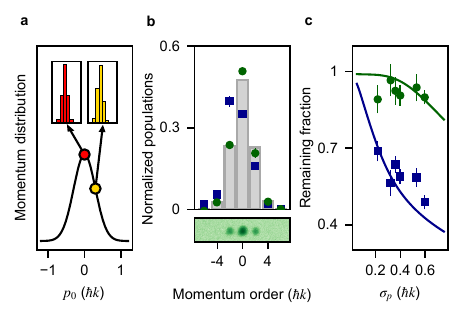}}
\caption{\textbf{Robustness of the Floquet state preparation.}
(a)~For a given acceleration sequence, the Floquet state depends on the initial atom momentum $p_0$. The insets show the Floquet states for $p_0 = 0\hbar k$ and $0.3\hbar k$.
(b)~Histogram of the momentum state distribution of the theoretical Floquet states $\ket{w_0(p_0)}$ averaged over the momentum distribution $f(p_0)$ and experimentally prepared Floquet states. Green circles (resp. blue squares) correspond to the state obtained with the robust preparation (resp. non-robust). Lower panel: Image of the prepared Floquet state with the robust preparation.
(c)~Remaining fraction of accelerated atoms at $20\hbar k$ for robust Floquet state preparation in $\ket{w_0}$ (green circles) and without robust optimization (blue squares) as a function of the momentum dispersion $\sigma_p$ of the atomic cloud.
Error bars are a statistical standard error of the mean over 20 realizations. The above examples correspond to an acceleration sequence based on tanh pulses of $8~\mu$s.}
\label{fig:Robustness}
\end{figure}

We have introduced the principle of Floquet acceleration for a pure initial momentum state $\ket{p_0}$. The OC pulses corresponding to this idealized situation, are called non-robust control. However, each $p_0$ within a BEC momentum distribution $f(p_0)$ is associated with a specific Floquet state $\ket{w_0(p_0)}$. For example, Fig.~\ref{fig:Robustness}(a) shows the decomposition into momentum states of two Floquet states associated with $0\hbar k$ and $0.3\hbar k$. Thus, the non-robust control does not achieve perfect preparation for the entire momentum distribution. Therefore, to efficiently implement this approach, we need to find a robust OC pulse capable of simultaneously preparing Floquet states $\ket{w_0(p_0)}$ for all $p_0$.

For acceleration sequences based on short $\pi$ pulses, the Floquet state is only slightly changed with respect to $p_0$, and robust OC pulses can be designed for a reasonably broad momentum distribution (see Methods). In Fig.~\ref{fig:Robustness}(b), we show in an illustrative example that the distribution measured in the momentum state basis with the robust optimal control is very close to the theoretical distribution, while significant differences are observed in the case of the non-robust control. The quality of the robust preparation is confirmed by the measurements of the fraction of atoms accelerated to $20 \hbar k$, as shown in Fig.~\ref{fig:Robustness}(c).

Our experimental results are in a good agreement with simulations performed without adjustable parameters for both cases. The acceleration efficiency achieved with the non-robust preparation shows a rapid decrease with increasing momentum dispersion, in contrast to the robust case, which maintains a high efficiency up to 0.35$\hbar k$ ($\sim 45 $~nK). Beyond this velocity dispersion, the OCT algorithm does not converge to robust control within the 100~$\mu$s duration of the OC-1 and OC-2 pulses.

\subsection*{Floquet atom acceleration}
\begin{figure}
\centerline{\includegraphics[width=0.5\textwidth]{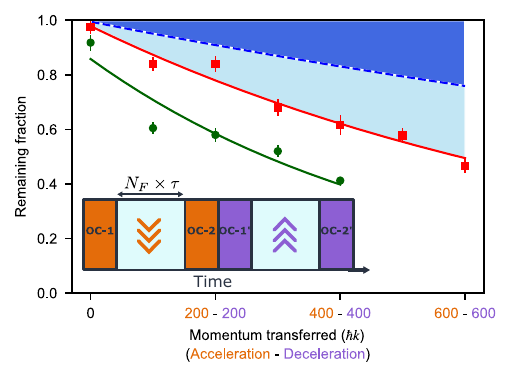}}
\caption{\textbf{Floquet Accelerator momentum transfer.} 
Fraction of remaining atoms after up to 600$\hbar k$ acceleration and 600$\hbar k$ deceleration. The red squares (resp. green circles) correspond to a square (resp. tanh) pulse sequence. Error bars are a statistical error on the mean over 20 realizations. Solid lines show fits to the data with the function $P_0 \cdot P^{2N_F}$, and the shaded areas indicate numerically estimated losses due to spontaneous emission (dark blue) and pulse-to-pulse amplitude fluctuations with a $4\%$ standard deviation (light blue). 
Inset: Scheme of the acceleration-deceleration sequence of $2\times N_F$-pulses showing the OC preparation sequences.
}
\label{fig:accelerations}
\end{figure}

We measure the efficiency of the Floquet accelerator using a sequence of acceleration-deceleration over a range of momentum values, up to an acceleration-deceleration of 600-600$\hbar k$ (i.e. a total transfer of 1200$\hbar k$). Figure \ref{fig:accelerations} shows the measured fraction of remaining atoms for two different acceleration sequences, one based on tanh pulses with a duration of $8~\mu$s and the other on square pulses with a duration of 5.3 $\mu$s. We fit these data with a function $P_0 \cdot P^{2N_F}$, where $N_F$ is the number of acceleration pulses. The parameter $P_0$ mainly represents the overall efficiency of the OC-1 and OC-2 stages and $P$ is an effective pulse-to-pulse efficiency of the acceleration process. We achieve an efficiency per $\hbar k$ ($\sqrt{P}$) of 0.99945(5) (resp. 0.9990(2)) for the square (resp. tanh) pulse sequence.

The observed efficiency is the highest reported to date~\cite{Kovachy12,Jaffe2018,McAlpine2020,Wilkason2022}. In addition, there is potential for improvement by mitigating spontaneous emission, a significant limitation as shown in Fig~\ref{fig:accelerations}, by increasing laser detuning and power~\cite{Kim20}. Another important factor to consider is amplitude fluctuations. Numerical simulations show that pulse-to-pulse fluctuations, characterized by a standard deviation of $4 \%$ (typical in our setup), can account for the observed efficiency. This problem can be mitigated by laser power stabilization techniques. This suggests that the intrinsic efficiency of Floquet acceleration allows the realization of large momentum transfers in regimes well above 1000$\hbar k$.

The demonstrated $2 \times 600\hbar k$ momentum transfer with an efficiency of 0.5 is achieved in 3.6~ms, including four OC pulses of $100~\mu$s each. Two of them are used here to reverse the direction of acceleration, allowing the atoms to return at a detectable velocity (see inset in Fig.~\ref{fig:accelerations}). This results in an average momentum transfer of $\hbar k$ every $3~\mu$s, the largest acceleration demonstrated so far with multi-photon transitions.

\subsection*{Large Momentum Transfer Atom Interferometer}
\begin{figure}
\centerline{\includegraphics[width=0.5\textwidth]{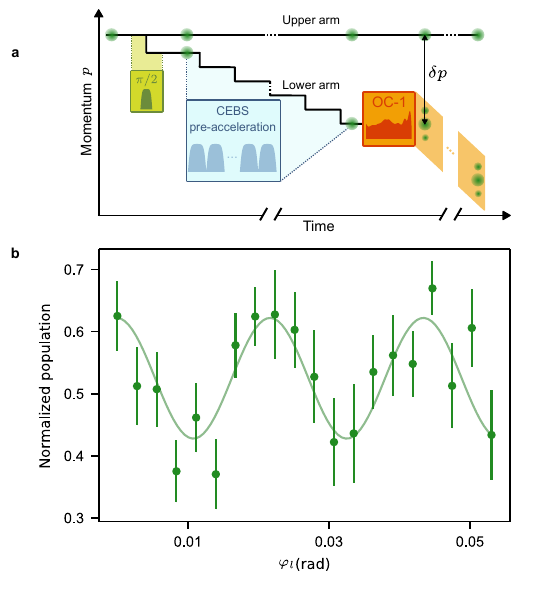}}
\caption{\textbf{Floquet accelerator-based atom interferometer.} 
(a)~Momentum-time diagram of the composite beam splitter: $\pi/2$ Bragg pulse, 10 CEBS pre-acceleration and 289 Floquet acceleration pulses leading to 600$\hbar k$. Green dots represent momentum states populated during the sequence. 
(b)~Fringes for 600$\hbar k$ LMT-interferometer. Each point is the averaged measurement over around 10 realizations and error bars are the standard errors on the mean. The solid line is a sinusoidal fit to the data.}
\label{fig:fringes}
\end{figure}

We implement the Floquet accelerator in an LMT inteferometer (see Fig.~\ref{fig:apparatus}(b)). The $\pi/2$ beam splitters used in the interferometer create two arms with a momentum separation of only 2$\hbar k$. To effectively implement Floquet accelerations in a LMT interferometer, it is essential to maintain decoupling between the acceleration processes in each arm. However, the Floquet state $\ket{w_0}$ is typically decomposed over many momentum states as can be seen, e.g. in Fig.~\ref{fig:Robustness}(b). Thus, during the early acceleration stage, when the momentum difference between the two interferometer arms is minimal, typically only a few $\hbar k$, the momentum components of the Floquet state of the accelerated arm would coincide with those of the unaccelerated arm. This overlap leads to a significant reduction of the LMT beam splitter efficiency.  To mitigate this effect, one needs to tailor an efficient initial two ports  beam splitter reaching a sufficient momentum separation such as high order Bragg diffraction~\cite{Beguin2022,saywell2023}, Bloch acceleration~\cite{Clade2009}, shaken-lattice beamsplitters \cite{Weidner2018}, or additional OC pulses~\cite{Goerz2023, Louie2023}. We choose to implement this pre-acceleration strategy (see Fig.~\ref{fig:fringes}(a)) using Coherent Enhancement of Bragg Sequences (CEBS)~\cite{Beguin2023} until the momentum difference $\delta p$ between the interferometer arms reaches 22$\hbar k$, corresponding to the combined momentum transferred by the $\pi/2$ pulse and the 10 CEBS pulses. The latter pulses used in the pre-acceleration stage have a duration of 40 $\mu$s, ensuring that only a very limited number of momentum states are populated during each pulse. We then switch to Floquet acceleration for the remainder of the acceleration sequence. Therefore, the maximum momentum separation is $2N\hbar k=(22+2 N_F)\hbar k$.

The interferometer signal is the normalized atom numbers detected in the two main output ports: $S~=~N_0/(N_2+N_0)$, where $N_2$ is the number of atoms measured in output port $-2 \hbar k$, and $N_0$ is associated with port $0 \hbar k$. The interference fringe between the two arms is described by a sinusoidal function $S(\phi)~=A(1+V\sin(\phi))$, where $V$ is the fringe visibility and $A$ is the mean value. The phase shift $\phi$ can be varied with the phase of the optical lattice, which is imprinted on the accelerated atomic wavefunction at each pulse transferring a momentum $2\hbar k$. In practice, an offset $\varphi_l$ of the optical lattice phase is applied during the $N_F$ Floquet acceleration pulses, resulting in a phase shift $\phi=N_F \times \varphi_l$. The fringe is scanned by incrementing $\varphi_l$. The visibility is determined by fitting the amplitude of the sinusoidal function.
The scaling with $N_F$ provides a compelling evidence that the interference is specific to the fully separated interferometer and not to signals associated with parasitic interferometers (see Methods). Thus, we demonstrate an atom interferometer with a separation of 600$\hbar k$ and a visibility of 18~$\% \pm 4\%$ (see Fig.~\ref{fig:fringes}(b)). This achievement represents the largest momentum separation ever achieved in an atom interferometer. The visibility value is mainly limited by atoms undergoing spontaneous emission reaching the detection volume and by background noise within the detection system. In addition, the maximum momentum separation results from the inherent time-of-flight constraints of our experimental setup (see inset of Fig.~\ref{fig:apparatus}(d)).

\subsection*{Discussion}
In conclusion, our work introduces a novel approach that combines Floquet formalism and optimal control protocols to accelerate atoms in optical lattices, covering both continuous and discrete acceleration scenarios. This method operates in a highly non-adiabatic regime, allowing accurate control of atomic phases. In particular, we identify a remarkable regime characterized by a constant amplitude and a discrete lattice frequency evolution. Our approach achieves unprecedented efficiency and speed in lattice-based acceleration, resulting in the highest and fastest demonstrated maximum momentum separation of 600$\hbar k$ in atom interferometry. The momentum separation is limited only by the dimensions of our vacuum chamber.

Despite this already high efficiency, a major limitation of our setup, spontaneous emission, could be overcome by increasing the detuning with the excited state, while maintaining the same lattice depth with more laser power. In addition, we show that the Floquet state can be accurately approximated by a squeezed state. This simple parameterization could allow refinement of the acceleration sequence, leading to improved robustness of matter-wave beam splitters, especially with respect to laser power fluctuations and a fine control of systematic effects associated with the atomic diffraction phase shifts. Therefore, this method has the potential to support interferometers well beyond $1000\hbar k$, opening new avenues for future applications in precision metrology, quantum technologies, and addressing one of the critical challenges for very large scale atom interferometers envisioned for gravitational wave detection. 

These results not only advance applications in atom interferometry but also introduce an innovative approach that extends the capabilities of quantum optimal control to navigate high dimensional Hilbert spaces~\cite{kochroadmap,Larrouy2020}. In our approach, we encapsulate the complexity within a Floquet state, allowing both remarkably fast and robust state-to-state preparation within a vast 300-dimensional Hilbert space. The ability to precisely manipulate quantum states in such complex systems has great potential for quantum technologies, either for sensing or for quantum computing and simulation~\cite{Frank2016,Larrouy2020,Dupont2023}. Therefore, beyond applications in atom interferometry, our approach provides an effective method in the quantum control toolbox.



\section*{Acknowledgments}
We acknowledge C. Salomon, B. Peaudecerf and J. Vigué for valuable discussions. This research was supported by the research funding grants QuCoBEC (ANR-22-CE47-0008), TANAI (ANR-19-CE47-000), and QAFCA “Plan France 2030” (ANR-22-PETQ-0005). SB acknowledges the support of the QuanTEdu-France program (ANR-22-CMAS-0001).


\section*{Contribution}
BA and AG conceived the project. TR, LC, and SB performed the experiment and data analysis. AB and TR participated in the construction of the apparatus and performed preliminary studies on sequential LMT. ED performed theoretical calculations under the guidance of DS. DGO, BA, DS, and AG contributed to the interpretation of the results, with AG leading the project. All authors participated in the preparation of the manuscript.


\clearpage


\section*{Methods}

\subsection{Quasi-Bragg Diffraction in a Nutshell}
The optical lattice consists of two counter-propagating beams, characterized in the laboratory reference frame by their frequencies ($\omega_{1,2}$), their opposite wave vectors ($\mathbf{k}_{1,2}$) with $\mathbf{k}_1 \sim - \mathbf{k}_2$, and a phase ($\varphi_{1,2}$). The phase and frequency differences between these two beams are denoted by $\varphi=\varphi_1-\varphi_2$ and $\omega_l = \dot{\varphi}=\omega_1 - \omega_2$, and the mean wave vector is defined as $k = (k_1 + k_2)/2$. When these two waves are superimposed, they form a quasi-stationary wave moving with a velocity $v = \omega_l / 2k$ relative to the laboratory reference frame.

The laser is detuned far from the frequencies of the atomic transitions, allowing for adiabatic elimination of the excited state. The atom-light interaction is then reduced to a light shift proportional to the light intensity. This leads to an interaction potential of the form $2\hbar \Omega(t) \sin^2(k z - \varphi(t)/2)$, where $\Omega(t)$ represents the two-photon Rabi frequency. The Hamiltonian describing the evolution of the atom is the sum of a kinetic energy term, and the potential associated with the standing wave (see~Supp. Mat.)
\begin{equation*}
\label{eq-H-refBragg}
    \hat{H}_2=\frac{1}{2M}\bigg(\hat{p}-\frac{M \omega_l}{2 k}-Mgt\bigg)^2 -\frac{\hbar \Omega(t)}{2} \left( e^{2 i k \hat{z}} + e^{-2 i k \hat{z}} \right).
\end{equation*}

The operators $e^{\pm 2 i k \hat{z}}$ couple momentum states differing by $2\hbar k$. The periodic potential can thus be interpreted as a two-photon process in which a photon is absorbed in one traveling wave and re-emitted by stimulated emission in the other wave, resulting in a $2\hbar k$ momentum transfer. Energy conservation leads to the two-photon Bragg resonance condition: $\omega_l = 4\omega_r + 2 kv_a$, where $v_a$ is the projection of the atomic velocity onto the lattice direction in the laboratory reference frame. Here, the optical lattice is vertical and the atoms are in free fall. Consequently, a time-dependent frequency ramp $\omega_l(t) = 4 \omega_r + 2k v_0 - 2kgt = \omega - 2kgt$ is set to compensate for the acceleration due to gravity and to maintain the Bragg condition, $\omega$ is the lattice frequency in the free fall frame.

For weak optical lattice depths, the dynamics of the system is described by an "effective two-level system" (Bragg approximation). Under these conditions, we observe Rabi oscillations between two momentum states separated by $2\hbar k$. The Rabi phase is defined as $\Theta_R=\int \Omega(t) \mathrm{d}t$, for $\Theta_R = \pi/2$ (resp. $\pi$) the corresponding pulse is called a $\pi/2$ pulse (resp. $\pi$ pulse). In the Bragg regime, a $\pi/2$ pulse creates an equiprobable coherent superposition similar to a beam splitter for the atomic wave function. A $\pi$ pulse reverses the two momentum states, essentially acting as a mirror for the atoms. For larger lattice depths, the Bragg approximation breaks down, requiring consideration of more complex dynamics between momentum states~\cite{Beguin2022}.

\subsection{Experimental setup}
\subsubsection*{Cold atom source}
The atomic source consists of an ensemble of Rubidium-87 atoms cooled by forced evaporation in an all-optical trap. The configuration of this dipole trap is based on two horizontally crossing beams at 1070~nm, plus a third beam at 1560~nm at an angle of $45^{\circ}$ to the vertical, with a smaller waist. This configuration allows the trapping frequencies and trapping depth to be adjusted independently, and the runaway regime to be achieved during evaporative cooling. In addition, a horizontal magnetic field gradient is applied during the evaporative cooling process to prepare the condensate in the pure $\ket{F=1,m_F=0}$ state. In 6 seconds we produce a Bose-Einstein condensate (BEC) consisting of $6 \times 10^4$~atoms. Confinement frequencies at the end of evaporation reach about $(60 \times 900 \times 1100)~\mathrm{Hz}^3$. By transferring the BEC to a less confining trap, characterized by frequencies around $(10 \times 80 \times 80)~\mathrm{Hz}^3$, we achieve a significant reduction in the velocity dispersion. This allows us to obtain atomic ensembles of $3\times 10^4$ atoms with a momentum dispersion of about $0.3\hbar k$ (corresponding to an effective temperature of 30~nK).

For the data presented in Fig.~\ref{fig:Robustness}(c), we adjust the momentum distribution either by adding a delta kick collimation step (for the data below 0.3$\hbar k$) or by changing the trap parameters during the final evaporative cooling steps.

\subsubsection*{Optical lattice}
Our 780 nm-optical lattice is made by frequency doubling of a 1560 nm-laser. The resulting 780 nm-beam is split into two to form the arms of a standing light wave used to create the optical lattice. The phase and frequency of each arm are controlled using a acousto-optic modulator in a double-pass configuration. The two beams are then recombined with orthogonal linear polarizations and coupled through a fiber. An additional acousto-optic modulator at the fiber output controls the amplitude of the two beams. The vertically aligned beams are retroreflected to create a standing light wave that forms the optical lattice. This retroreflected configuration implements a double lattice with opposite effective wave vectors and orthogonal circular polarizations, as described in detail in~\cite{Beguin2022}. The Bragg transitions for both lattices are degenerate for a vanishing relative atom-lattice velocity. However, a brief free fall induces a Doppler shift that causes one lattice to go out of resonance. Each lattice operates at an approximate power of 250~mW, with a fixed detuning of 40~GHz from the atomic resonance. At the atom positions, the waist measures approximately 1.6~mm, resulting in a maximum lattice depth characterized by the peak two-photon Rabi frequency $\Omega_{\mathrm{max}}= 25 \omega_r$.

The relative atom-lattice velocity is set by the frequency difference between the two beams that form the lattice~\cite{Beguin2022}. A frequency chirp is systematically added to the frequency difference profile to compensate for the atomic free fall and create a quasi-standing wave in the free fall frame. 

\subsubsection*{Detection}
Atoms are measured through fluorescence imaging after time-of-flight. Given the typical velocity dispersion of the atomic cloud, the final free fall time must be larger than 14~ms to separate the two adjacent momentum states distant from $2\hbar k$. The fluorescence beams are the laser beams used for initial laser cooling brought at resonance. The beam waist of 7~mm, and the limited field of view of the camera limit the detection volume to a typical size of less than 1~cm and thus the overall available time of free-fall to $\sim 45$~ms.

For all the data, the population in each momentum state is measured as the fluorescence signal integrated over a square box, typically $60~\mu$m of size, centered on the center of mass position of the considered state. For the data presented in Fig.~\ref{fig:Robustness} and \ref{fig:accelerations}, the populations are normalized to the total atom number of a free fall BEC at the same position on the camera without any interaction with the optical lattice. For Fig.~\ref{fig:fringes}, data are normalized to the sum of the populations in the two main output states.

\subsection{Visibility measurements and analysis}
Interferometer fringe data (see Fig.~\ref{fig:fringes}) are fitted with the following function: $\frac{N_0}{N_0+N_2} =A\left( 1+V \sin{(K \varphi_l + \varphi_0)}\right)$, where the offset $A$, the visibility $V$ and the offset phase $\varphi_0$ are the three fitted parameters. $K$ is a known scaling factor with the phase jump $\varphi_l$ (see below).

For this measurement procedure, the positions of the two boxes measuring the populations in the two output ports are slightly adjusted within the cloud size to optimize the signal-to-noise ratio of the fitted visibility. The position changes have a marginal effect on the detected atom number or on the value of the fitted parameter (visibility and offset phase).

The statistical significance of the fitted visibility has been checked. We have simulated a large number ($N_s = 10^6$) of fully random datasets (with exactly zero visibility) each with the same size as the measurements represented in Fig.~\ref{fig:fringes}(b). The variance of the normal distribution corresponds to the observed detection noise. Each dataset is then analyzed with the same procedure for the presented measurement from which we construct a number $V/\sigma_V$ where $V$ is the fitted visibility and $\sigma_V$ the corresponding uncertainty. 
From the histogram of the $N_s$ values of $V/\sigma_V$, we compute the complementary cumulative distribution function. 
For a given value $x$, this function is the probability to obtain a fitted $V/\sigma_V$ larger or equal to $x$ on a random dataset. Evaluating the complementary cumulative distribution function of the measured value of $V/\sigma_V$ (\textit{i.e.} $x=4.8$) gives the probability to fit a value at least as extreme from a dataset with zero visibility.
For the presented measurement, we estimate this probability to be less than $10^{-5}$, which significantly rejects the no visibility hypothesis.

\begin{figure}
\centerline{\includegraphics[width=0.5\textwidth]{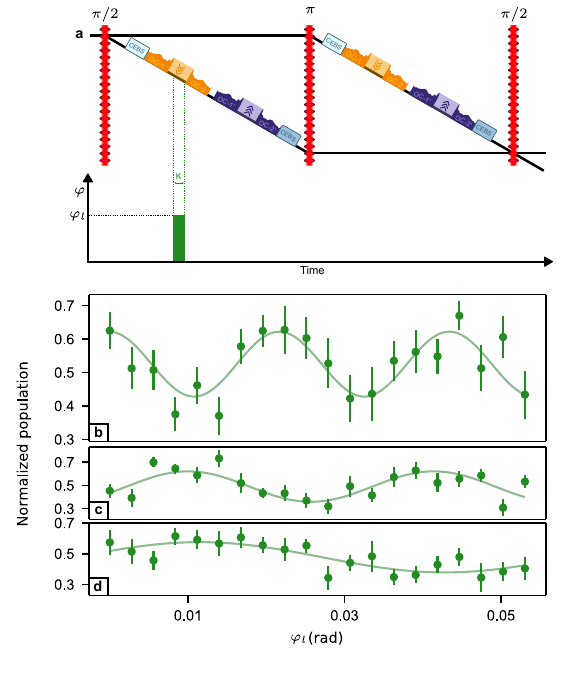}}
\caption{\textbf{Extended data: Floquet accelerator-based atom interferometer.} 
(a)~Schematic representation of the interferometer sequence. The phase jump $\varphi_l$ is applied during the $K$ last Floquet acceleration pulses of the first acceleration sequence. 
(b)-(d)~Fringes for a maximum momentum separation of $600\hbar k$ for three different scaling of the phase. Dots are mean values of the normalized population over around 10 realizations and the error bars are the corresponding standard errors on the mean. (b)~The phase jump $\varphi_l$ is applied during all the Floquet acceleration pulses of the first acceleration sequence ($K=289$) . (c) and (d)~Same as (b) but with $K=200$ and $K=100$, respectively. 
}
\label{fig:ExtendedFringes}
\end{figure}

To further strengthen the visibility measurement, we performed experiments with the same maximum momentum separation $2N\hbar k = (22+2N_F)\hbar k$ but changing the scaling factor $K$. It corresponds to the number of acceleration pulses that experience the phase jump $\varphi_l$ among the $N_F$ pulses of the acceleration sequence (see extended data Fig.~\ref{fig:ExtendedFringes}(a)). Figure~\ref{fig:ExtendedFringes}(b) shows the same data presented in Fig.~\ref{fig:fringes}(b) with $K = N_F = 289$. Fringes for $K=200$ and $K=100$ are also given in Fig.~\ref{fig:ExtendedFringes}(c) and (d) respectively. The three measurements show a sinusoidal behavior with the expected scaling with the phase jump and consistent fitted visibilities ($V=18\pm4\%$ for $K= 289$, $V=27\pm5\%$ for $K=200$ and $V=21\pm4\%$ for $K=100$). It confirms that the visibility measurement is not biased by parasitic interferometers at this level of uncertainty.

\subsection{Interferometer sequence}

The full interferometer sequence is illustrated in Fig.~\ref{fig:ExtendedFringes}(a). It starts with a first beam-splitter creating the coherent superposition between momentum states $\ket{p_0}$ and $\ket{p_0-2\hbar k}$. The lower arm is further accelerated by a serie of pre-acceleration pulses using the CEBS. The state is then sufficiently separated from $\ket{p_0}$ (upper arm) to be transformed into the Floquet state by the preparation pulse OC-1 and accelerated by a serie of $N_F$ pulses. A second preparation pulse OC-2 transforms it back to the fully accelerated state. A free evolution time $T'$ is added before the symmetric deceleration sequence. In practice, for the experiments reported in this paper, $T'$ remains small (below 1~ms). After a central miror pulse exchanging the momentum states, the interferometer is closed by a symmetric sequence addressing the upper arm and a final beam splitter.

\subsubsection*{Beam splitter, mirror and pre-acceleration pulses}

Except for the OC preparation sequences and Floquet-acceleration pulses, for which the shape are discussed in the dedicated sections, all the pulses use hyperbolic tangent amplitude profiles $\Omega(t) = \Omega_0 \times \mathrm{max}\left\{0,\tanh\left(8t/\tau_0\right)\tanh\left(8(1-t/\tau_0)\right)\right\}$ where $\tau_0$ is the total duration of the pulse and $\Omega_0$ the peak Rabi frequency. In particular, the beam splitter (resp. mirror) pulses have a duration $\tau_0 = 50~\mu$s (resp. $\tau_0 = 60~\mu$s) and the corresponding amplitude $\Omega_0$ to produce the expected Rabi phase when coupling momentum states $\ket{0}$ and $\ket{-2\hbar k}$ at the two-photon resonance. Similarly, the pre-acceleration pulses using the CEBS technique are mirror pulses with the same amplitude profile of duration $40~\mu$s following the accelerated trajectory. 

\subsubsection*{Robust optimal control}
We use optimal control theory~\cite{PRXQuantumsugny} and a gradient-based algorithm, GRAPE~\cite{khaneja_optimal_2005}, to design the control pulses OC-1 and OC-2. The optimal control problem is to maximize the figure of merit $F_1$ defined as
\begin{equation*}
F_1=\int_{-\infty}^{+\infty}|\langle\psi(\tau_c)|w_0(p_0)\rangle|^2f(p_0)\mathrm{d}p_0,
\end{equation*}
where $|\psi(t)\rangle$ is the solution of the Schr\"odinger equation at time $t$ in the free fall frame for a given value of the parameter $p_0$ and $\tau_c$ the duration of the optimal control. Note that one could define a different figure of merit that takes into account the phase of $\ket{w_0(p_0)}$ (see Supp. Mat.). The algorithm optimizes both the amplitude and the frequency of the optical lattice over time to achieve the expected target fidelity. The corresponding control solution is said to be robust to $p_0$ in the sense that the same control protocol is used for all values of this parameter in the range $[-3\sigma_p,3\sigma_p]$. Details about the numerical implementation can be found in the Supp. Mat.

In this study, we considered a given acceleration sequence corresponding to a Floquet state $\ket{w_0}$ and transformed the input state to this Floquet state using optimal control protocols. This solution is the most efficient and provides the highest momentum transfer rate. However, there is another optimization strategy based on Floquet analysis, which is to apply optimal control algorithms to design pulses (amplitude and frequency) of the acceleration sequence itself. This pulse shaping aims to generate a Floquet state with a significant overlap with the initial momentum state $\ket{p_0} \approx \ket{w_0}$. This strategy has resulted in a less efficient and much longer acceleration sequence.

\begin{figure}
\centerline{\includegraphics[width=0.5\textwidth]{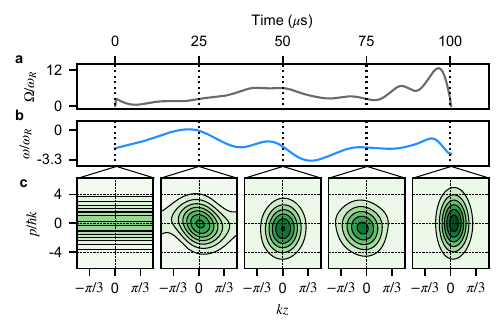}}
\caption{\textbf{Extended data: Optimal control profiles for Floquet state preparation.} 
(a)-(b)~Amplitude and frequency profiles used to prepare the Floquet state $\ket{w_0}$ for a square pulse sequence of period $\tau=5.3~\mu$s.
(c)~Evolution of the state during the OC preparation sequence in the Husimi representation. The initial state (for $t=0$) is the plane wave $\ket{p_0}$. The target Floquet state $\ket{w_0}$ is obtained at $t=100~\mu$s and is very similar to a displaced squeezed state in phase space.
}
\label{fig:OCTexample}
\end{figure}
\subsubsection*{Example of OC preparation sequence}
Figure~\ref{fig:OCTexample} gives an example of the OC pulse designed for a Floquet acceleration based on square amplitude with a duration $\tau =5.3~\mu$s. The total duration of the OC preparation sequence is constrained to 100~$\mu$s. Extensive numerical simulations show that this time is a good compromise between efficiency of the process and the relatively simple shape of the optimal pulse. The time evolution of the state during the OC pulse is illustrated in Fig.~\ref{fig:OCTexample}(c). It is continuously transformed from a non-localized initial plane wave $\ket{p_0}$ to the target Floquet state $\ket{w_0}$, similar to a displaced squeezed state in phase space. Note that the reverse OC sequences used in this work (labeled OC-2 in the main text) have a similar global shape and are close to the time-reversal of the profiles displayed in Fig.~\ref{fig:OCTexample}. 



%


\end{document}